\documentclass{article}

\usepackage{nips_2017}

\usepackage[utf8]{inputenc} 
\usepackage[T1]{fontenc}    
\usepackage{hyperref}       
\usepackage{url}            
\usepackage{booktabs}       
\usepackage{amsfonts}       
\usepackage{nicefrac}       
\usepackage{microtype}      
\usepackage[utf8]{inputenc}
\usepackage{graphicx}
\usepackage{subcaption}
\usepackage{amsmath}

\title{Rapid point-of-care Hemoglobin measurement through low-cost optics and Convolutional Neural Network based validation}

  \author{
  Chris Wu\\
  Athelas Inc. \\
  \texttt{chris@athelas.com}
  \And
  Tanay Tandon\\
  Athelas Inc. \\
  \texttt{tanay@athelas.com}
}
\begin{document}
\maketitle
\begin{abstract}
  A low-cost, robust, and simple mechanism to measure hemoglobin would play a critical role in the modern health infrastructure. Consistent sample acquisition has been a long-standing technical hurdle for photometer-based portable hemoglobin detectors which rely on micro cuvettes and dry chemistry. Any particulates (e.g. intact red blood cells (RBCs), microbubbles, etc.) in a cuvette's sensing area drastically impact optical absorption profile, and commercial hemoglobinometers lack the ability to automatically detect faulty samples. We present the ground-up development of a portable, low-cost and open platform with equivalent accuracy to medical-grade devices, with the addition of CNN-based image processing for rapid sample viability prechecks. The developed platform has demonstrated precision to the nearest $0.18[g/dL]$ of hemoglobin, an \(R^{2} = 0.945\) correlation to hemoglobin absorption curves reported in literature, and a 97\% detection accuracy of poorly-prepared samples. We see the developed hemoglobin device/ML platform having massive implications in rural medicine, and consider it an excellent springboard for robust deep learning optical spectroscopy: a currently untapped source of data for detection of countless analytes.
\end{abstract}
\section{Introduction}
Hemoglobin is one of the most common blood tests requested by clinics, and can be used in conjunction with other metrics to diagnose a host of diseases and conditions \cite{mayoclinic}\cite{biomed}, quantify the effects of pharmaceutical drugs \cite{medlineplus2}, and provide a holistic health benchmark \cite{biomed}. As roughly a quarter of the world’s population suffers from a form of hemoglobin deficiency \cite{who}, there are many niches where accessible hemoglobin measurement would fulfill significant unmet need \cite{cdc}\cite{drugsaging}\cite{biomed}.

The widely-adopted approach to point-of-care hemoglobin measurement involves hemolysing whole blood and converting hemoglobin derivatives for single-wavelength absorption measurement, followed by a simple Beer Law calculation \cite{hemocue}\cite{vanzetti}\cite{oshiro}. In present hemoglobin monitors the single biggest risk of inaccurate counting and as a result, incorrect clinical decision making, is incomplete hemolysis due to variable reagent performance. Consequently, light scattering interference from intact cells (lipid membranes) have a drastic effect on spectrophotometric output \cite{lipemia}.

Our intention in developing the presented device platform was twofold: to create an effective and inexpensive hemoglobin solution without the need for hazardous dry reagents such as cyanide (Drabkin's method) or sodium azide \cite{hemocue}\cite{vanzetti}\cite{azide}, and to demonstrate the efficacy of convolutional neural network based validation for elimination of a long-standing error source in hemoglobin measurement. 

\section{Detection platform summary}

The current version of the device has demonstrated: 1) precision to the nearest $0.18[g/dL]$ of Hb, 2) stability of measurement over time, with no fluctuation from a single sample over 10 minutes of continuous output, and 3) \(R^2 = 0.945\) correlation to optical transmission-Hb curve reported in literature (Fig.\ref{fig:HbCorr} image A).

\subsection{Device platform}

\begin{figure*}[!ht]
\centering
    \includegraphics[width=0.8\textwidth]{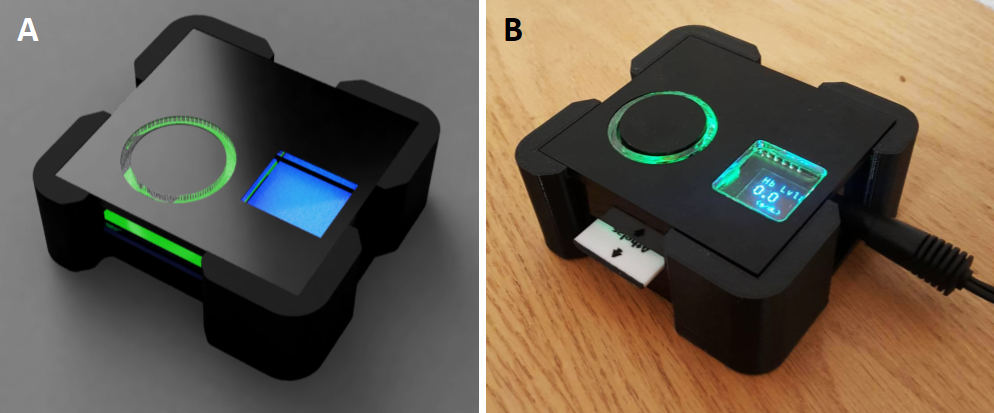}
    \caption{Image A) Design concept rendering. Image B) Completed device with blank strip and power adapter inserted.}
    \label{fig:Chassis}
\end{figure*}

\subsubsection{Sensor configuration}
For Hb measurement, a paired $540[nm]$ emitter-detector setup is used for sample transmission calculation. The LED and photodiode were selected to maximize light emission, photocurrent, and Hb absorption.

Image pre-processing for sample viability assurance are captured with a Pi Camera attached to a low-cost optical microscopy setup (separate from device at this point in development). See figure 4 for optical microscopy imaging examples. Both setups have a test strip insertion slot and a plane for allowing optical analysis.  
 
\subsubsection{Signal processing}
An $LTC1050$ Chopper-stabilized op amp is used in a standard transimpedance amplifier configuration for signal processing with a reverse-biased photodiode for signal linearity. Additional components protect sensitive analog signals from various forms of electromagnetic interference, including high frequency noise and LC-tank oscillation due to inherent component properties. As absorption is taken at steady-state, signal bandwidth is not a design concern.

Hb concentration is linearly proportional to optical absorbance ($A$), which is calculated from the negative log of fractional transmission, $Tf/Ti$. From Beer’s Law:
\begin{equation}
C = -Klog(\frac{Tf}{Ti})
\end{equation}
for $K = 1/(el)$, where $e$ is the molar extinction coefficient of hemoglobin, $l$ is the transmission path length ($100[\mu m]$), and $C$ is the concentration of hemoglobin in the sample (in $[g/dL]$).

The device's microcontroller contains a 10-bit ADC with a minimum voltage increment of $5[mV]$. With a swing of $1.5[V]$ the resolution (i.e. the smallest detectable change in sample percent transmission) is 0.333\%. Readjusting $R1$ to maximize full output swing ($5[V]$) will improve resolution in the next device iteration to 0.1[\%T]. For clinically relevant Hb levels and 0.333[\%T] increment size, the resolution is $0.18 [g/dL]$ Hb  (i.e. the difference between consecutive Hb values over a 0.333[\%T] step). This exceeds the precision of predicate devices such as Hemocue, which has an overall bias $\pm$ stdev of -$0.1 \pm 1.6[g/dL]$ compared to lab-grade hematology analyzers \cite{shah}. Of course, this level of precision from the presented device assumes no variability in capillary strip performance: a hefty assumption. 

\begin{figure*}[!ht]
\centering
    \includegraphics[width=0.8\textwidth]{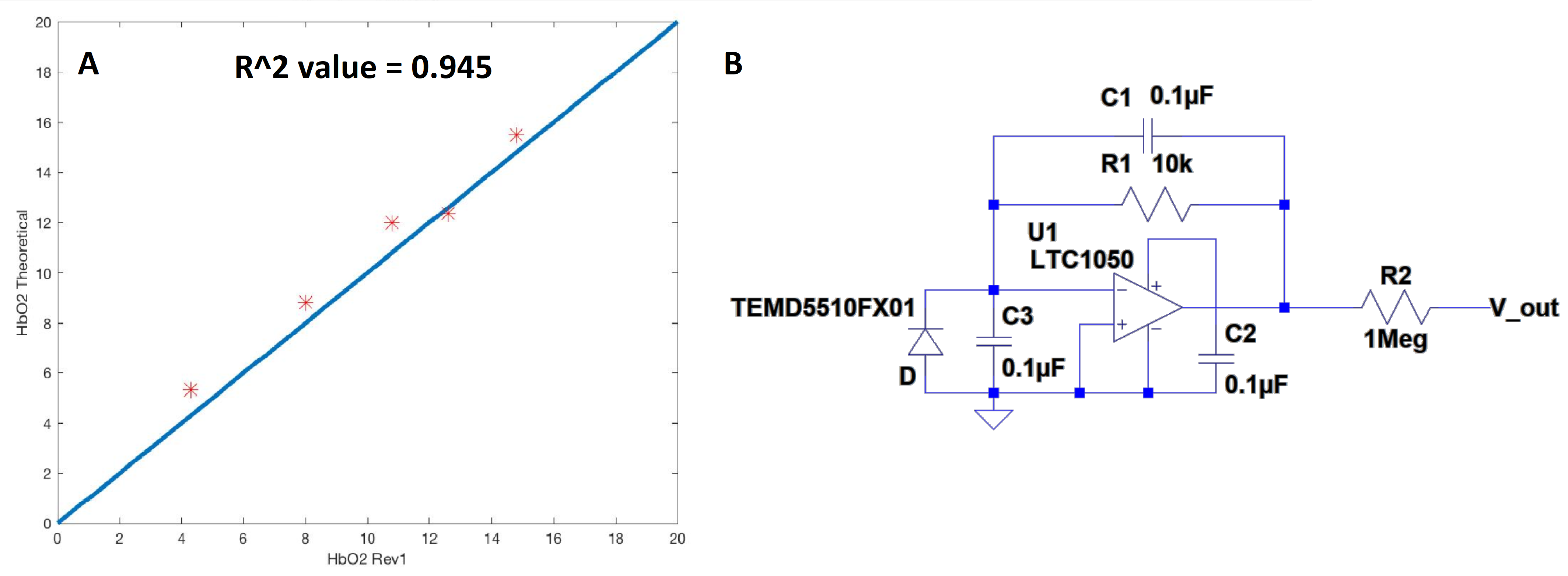}
    \caption{Image A) Plot of literature-reported Hb values for observed \%T measurements (extrapolated from $540[nm]$ molar extinction coefficient \cite{coeff1}\cite{coeff2}\cite{coeff3}) vs. the developed device readout for those \%T values. Strong correlation demonstrates hardware precision and accuracy over the full clinically-relevant range of hemoglobin levels. Image B) PIN photodiode and first amplifier stage used for signal acquisition.}
    \label{fig:HbCorr}
\end{figure*}
\subsection{Capillary test strip}

The test strip consists of a 100uM microchannel coated with sodium lauryl sulfate (SLS), a non-toxic surfactant which serves the dual purpose of lysing RBCs to eliminate turbidity and converting hemoglobin derivatives into a color-stable complex, SLS-Hb.
\begin{figure*}[!ht]
\centering
    \includegraphics[width=0.80\textwidth]{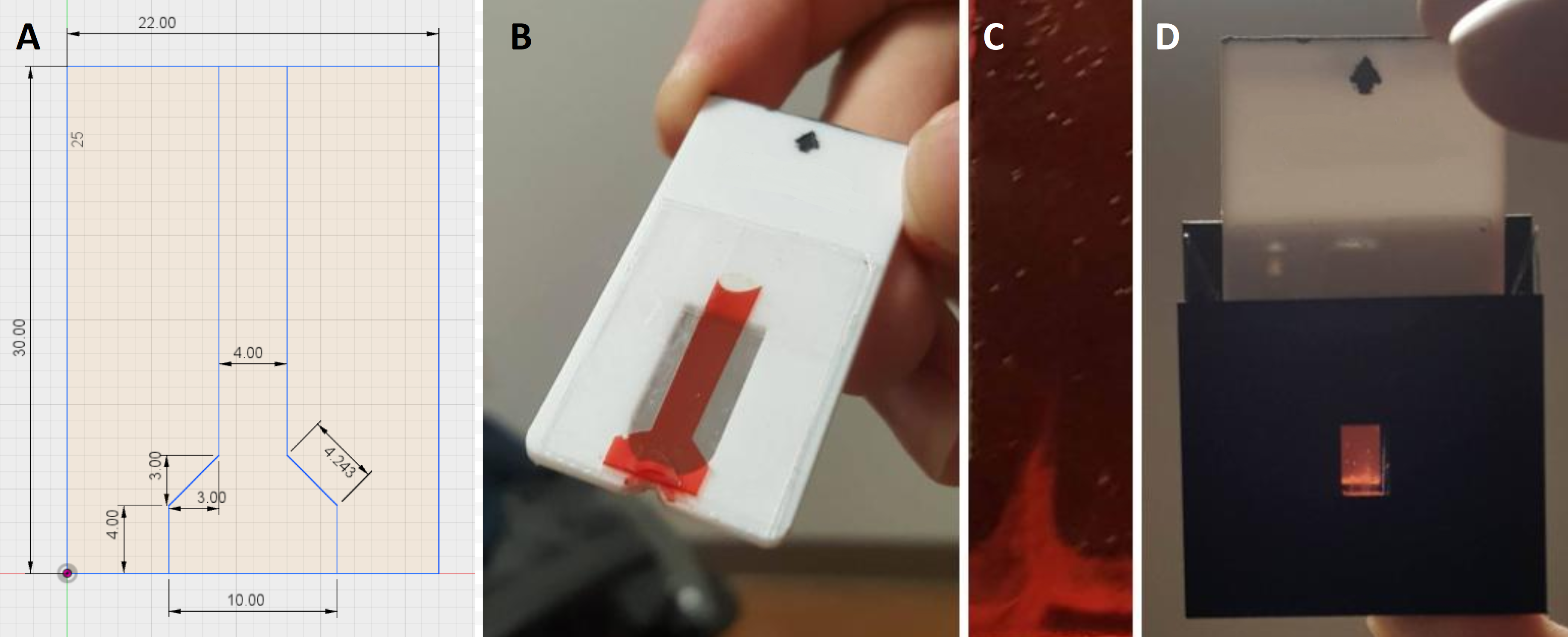}
    \caption{Image A is a diagram of the capillary strip design used. Strip with blood sample shown in B. Close-up of a channel demonstrating common issues with dry-reagent capillary strip methods in image C: turbidity at the bottom of the channel due to intact RBCs and air bubbles (white specks) at top. Image D shows the mechanical frame of the sensor input slot with the sample transmission measurement window.}
    \label{fig:blood}
\end{figure*}

In standard practice with SLS, $20[uL]$ of whole blood is added and mixed with $5[mL]$ of $2.08[mmol/L]$ SLS solution for a 0.52 moles SLS to blood sample volume ratio \cite{oshiro}. To adapt for dry-chemistry, $17[\mu mols]$ of SLS in aqueous solution were deposited and spread evenly onto each coverslip, for a total of $34[\mu mols]$ per $66[\mu L]$ available sample volume, maintaining the ratio.

However, since all fluid entering a capillary strip generally shares the same path, depending on viscosity and RBC density, the first units of blood entering the channel may use up reagent at the entrance, leaving none for subsequent units (hence incomplete hemolysis). This issue poses a critical error-source common to all dry capillary strip detection methods, which can be minimized through clever strip design.

A funnel-shaped channel (Fig.\ref{fig:blood} image A) was used to help mitigate the “use up” issue described. Blood deposited at the entrance of the funnel spreads out to cover a larger area before entering the narrow measurement area, having been hemolyzed by reagent in the funnel. While this design was generally effective in eliminating turbidity in the measurement window (Fig.\ref{fig:blood} panel D), variation in RBC density and sample viscosity prevented proper mixing/reaction on occasion, precluding this dry-chemistry approach from large-scale, frequent use with minimal user training. Only an automated viability check to auto detect such errors when present can completely eliminate this risk. 

\subsubsection{CNN-based image processing for sample viability check}

A core component of the presented system is the trained machine learning model for sample validation.

A separate imaging module was developed to analyze intact cells, air bubbles, and ensure appropriate sample prep. A filled strip is inserted into the imaging module, several images are taken across the strip at a scaled magnification, and a trained convolutional neural network determines the viability of the testing strip. 
\begin{figure*}[!ht]
\centering
    \includegraphics[width=1\textwidth]{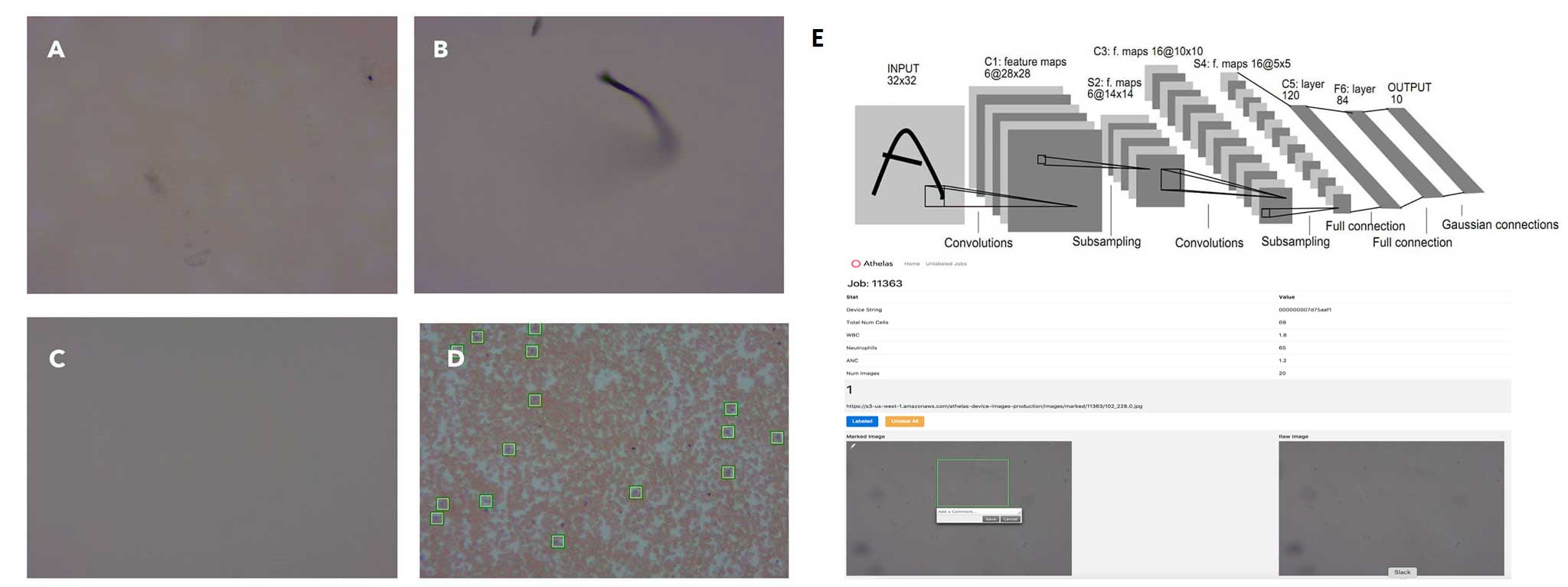}
    \caption{Images A,B,D) Image captures from the optical microscopy setup identified by the model as inadequate for spectrophotometry. Sample A shows debris and color instability. Sample B has image blur and particulates. Sample D contains intact cells which act as light scatterers. Additionally, green boxes identify white blood cells from the sample, demonstrating expandability of model for a vast number of use cases. Image C) Clear view with color stability and no scatterers: ideal for photometric hemoglobin measurement. Data was collected using a labeling interface of all run test strips (bottom of E). 200 strip images were labeled by a human identifying sections of debris and cell boundaries, and the overall image labeled "good" vs. "bad".}
    \label{fig:CNN}
\end{figure*}
The model was adapted from LeNet structure, and trained on 200 human labeled images of good/bad test strips. The images were augmented to expand the training set by 8x (rotational, translational, brightness, and hue transformations). These results were then cross validated on a test set of 100 strip images. 

The binary classification model performed with 97\% accuracy on the task across the non-augmented test set - this included filled strips by clinicians and general users. Similarly, cell segmentation modules were developed to threshold cell boundaries with each candidate cell being trained in a separate Convolutional Neural Net for type classification (White Blood Cell, Red Blood Cell, etc.). 
\section{Conclusion}
Our platform builds off decades of hemoglobin spectroscopy research, while providing unprecedented error detection and reliability thanks to convolutional neural networks. We believe this deep learning approach to on-board quality control can be rapidly scaled to other applications, boosting clinical performance, end diagnoses, and patient safety. 

Given the low-cost, open microcontroller nature of the presented device, we see the platform and model as a powerful combination for deploying in decentralized or rural areas. Untrained test operators have been a major reason point-of-care, self-administered hematology tests have remained unfeasible. However, with robust error correction, the risk of misdiagnosis is greatly reduced and this radical model of personalized medicine is brought within the realm of possibility.

With high-precision analog sensing and CNN based image processing, we have developed the foundation for multi-wavelength spectrophotometry, enabling both structural and molecular sample analyses. We are convinced that the confluence of these distinct, rich sources of data with the speed and versatility of AI will present numerous opportunities to advance the future of healthcare.

\section{Acknowledgements}
We are sincerely thankful to Dhruv Parthasarathy, Deepika Bodapati, Louis Virey, Steve Moffatt, and Sreevaths Kasireddy for funding acquisition and technical input in various aspects of this project.

\end{document}